\documentclass[twocolumn,pra,10pt,showpacs]{revtex4}
\usepackage{graphicx,amsmath,amssymb}

\newcommand{\beq}{\begin{equation}}
\newcommand{\eeq}{\end{equation}}
\newcommand{\beqa}{\begin{eqnarray}}
\newcommand{\eeqa}{\end{eqnarray}}
\newcommand{\ket}[1]{| #1 \rangle}
\newcommand{\bra}[1]{\langle #1 |}

\newcommand{\opdg}{\hat{\Delta}_{Q_u}}

\newcommand{\density}{\hat{\rho}}

\newcommand{\g}{\Gamma}

\newcommand{\gmax}{\Gamma_{{\rm \mbox{sup}}}}

\newcommand{\oprho}{\hat{\rho}}

\begin{document}


\title{Entanglement measure for general pure multipartite quantum states}

\author{Hoshang Heydari}
\email{hoshang@imit.kth.se} \homepage{http://www.ele.kth.se/QEO/}
\affiliation{Department of Microelectronics and Information
Technology, Royal Institute of Technology (KTH), Electrum 229,
SE-164 40 Kista, Sweden}

\author{Gunnar Bj\"{o}rk}
\affiliation{Department of Microelectronics and Information
Technology, Royal Institute of Technology (KTH), Electrum 229,
SE-164 40 Kista, Sweden}

\date{\today}

\begin{abstract}
We propose an explicit formula for an entanglement measure of pure
multipartite quantum states, then study a general pure tripartite
state in detail, and at end we give some simple but illustrative
examples on four-qubits and m-qubits states.
\end{abstract}


\maketitle

\section{Introduction}

One of the unsolved problems of modern quantum theory is the
quantification of  multipartite state entanglement
\cite{Lewen00,Dur99}. This is a task that is directly linked to
mathematics, such as linear algebra, geometry and functional
analysis. The definition of separability and entanglement of a
multipartite state was introduced in \cite{Vedral97,Vedral98}
following the definition for bipartite states, given in 1989 by
Werner \cite{Werner89}. Eventually, quantitative measures, such as
the entanglement of formation and concurrence were formulated for
bipartite systems \cite{Wootters}. In recent years, there have
been attempts to find an entanglement measure for qubit-qudit
states \cite{Rungta,Albeverio,Audenaert,Gerjuoy} and for
multipartite states, i.e., in
\cite{Plenio00,Bennett01,Dur00,Eisert01,Verst03,Eckert02}. To
exemplify, upper and lower bounds for the quantum relative entropy
of entanglement of a multipartite systems in terms of the
bipartite entanglements of formation, distillation, and quantum
entropy of various subsystems are derived in \cite{Plenio00}.
Measures based on the Schmidt rank are proposed in \cite{Eisert01}
and on local unitary and filtering processes in \cite{Verst03}.
Furthermore, in \cite{Hor00}, a very useful tool to detect
entanglement, called entanglement witness, is generalized to
multipartite states. The tool is a consequence of the Hahn-Banach
theorem which states that for any convex, compact, vector set
$\mathcal{S}$, if $\rho\notin\mathcal{S}$, there exists a
hyperplane that separates $\rho$ from $\mathcal{S}$. However, to
find such an operator, even in case of tripartite state, is a
formidable task. None-the-less, quite impressively, Ac\'{i}n
\textit{et al.} managed to construct a witness operator for a
class of mixed tripartite states \cite{Acin01}.

In a recent paper, Partovi proposes an entanglement
measure based on generalized Schmidt-decompositions of a state
\cite{Partovi}. In essence, his measure gives the difference
between the minimum entropy of the separable state having the same
statistical marginal distribution as the state being
characterized, and the entropy of the state itself. To find the
minimal difference, Partovi employs a successive
Schmidt-decomposition of the state. The measure quantifies the
logarithmic ``quantumness'' of a state irrespective of the type of
entanglement is possesses.

In this paper, we propose another measure of
entanglement for arbitrary, pure multipartite states.
Inspired by the work in \cite{Soto02}, we give an explicit
expression for such a functional. Our method is based on the joint
relative-phase properties of a multipartite quantum system
$\mathcal{Q}=\mathcal{Q}_{1}\otimes\mathcal{Q}_{2}\otimes\cdots\otimes\mathcal{Q}_{m}$
on a Hilbert space
$\mathcal{H}_{\mathcal{Q}}=\mathcal{H}_{\mathcal{Q}_{1}}\otimes\mathcal{H}_{\mathcal{Q}_{2}}
\otimes\cdots\otimes\mathcal{H}_{\mathcal{Q}_{m}}$ expressed by a
positive operator value measure (POVM) $\Delta_\mathcal{Q}$ on
$\mathcal{H}_{\mathcal{Q}}$. The POVM is constructed by taking the
$m$-fold tensor product of the subsystems' corresponding POVMs. We
have already discussed, in detail, our degree of entanglement for
a bipartite state in \cite{Hosh1,Hosh2}, so here we will only give
examples for multipartite states.

\section{Entanglement from a relative-phase POVM}

A general and symmetric POVM in a single $N_{u}$-dimensional
Hilbert space $\mathcal{H}_{\mathcal{Q}_{u}}$ is given by
\begin{equation}
\opdg = \sum^{N_{u}}_{l_{u}}\sum^{N_{u}}_{k_{u}=1}
e^{i\varphi_{k_{u},l_{u}}}\ket{k_{u}}\bra{l_{u}} ,
\end{equation}
where $\ket{k_{u}}$ are the basis vectors in
$\mathcal{H}_{\mathcal{Q}_u}$ and
\begin{equation} \varphi_{k_{u},l_{u}}=
-\varphi_{l_{u},k_{u}}(1-\delta_{k_{u} l_{u}}) . \label{eq:phase
relations}\end{equation} The POVM is a function of the
$N_{u}(N_{u}-1)/2$ relative phases
$(\varphi_{1_u,2_u},\ldots,\varphi_{1_u,N_u},\varphi_{2_u,3_u},\ldots,\varphi_{N_{u}-1,N_{u}})$.

It is now possible to form a POVM of a multipartite system by
simply forming the tensor product
\begin{eqnarray}
\nonumber
\hat{\Delta}_\mathcal{Q}(\varphi_{\mathcal{Q}_{1};k_{1},l_{1}},\ldots,\varphi_{\mathcal{Q}_{1};k_{m},l_{m}})=
& & \nonumber \\
\hat{\Delta}_{\mathcal{Q}_{1}}(\varphi_{\mathcal{Q}_{1};k_{1},l_{1}})
\otimes\cdots
\otimes\hat{\Delta}_{\mathcal{Q}_{m}}(\varphi_{\mathcal{Q}_{m};k_{m},l_{m}}),
&&
\end{eqnarray}
where, e.g., $\varphi_{\mathcal{Q}_{1};k_{1},l_{1}}$ is the set of
POVM relative phases associated with subsystems $\mathcal{Q}_{1}$,
for all $k_{1},l_{1}=1,2,\ldots,N_{1}$, where we need only to
consider when $l_{1}>k_{1}$ due to (\ref{eq:phase relations}). We
can now recast this POVM, expressed in local properties, in terms
of the relative-phase sums $\phi_{k_{1},l_{1},\ldots,k_{m},l_{m}}
= \sum_{u=1}^m \varphi_{k_{u},l_{u}}$. Note that if, e.g.,
$l_{v}=k_{v}$, then the term $\varphi_{k_{v},k_{v}}$ vanishes from
the sum due to (\ref{eq:phase relations}). From
$\hat{\Delta}_\mathcal{Q}$ we can form an associated real function
expressed in $\Pi_{u=1}^m N_u(N_u-1)/2=M$ linearly independent
relative-phase sums:
\begin{eqnarray}
\nonumber \mathcal{P}(\phi_{k_{1}^{(1)},l_{1}^{(1)},
\ldots,k_{m}^{(1)},l_{m}^{(1)}},\ldots,\phi_{k_{1}^{(M)},l_{1}^{(M)},
\ldots,k_{m}^{(M)},l_{m}^{(M)}})  = & &\nonumber \\
\mathrm{Tr}\left(\hat{\rho}\hat{\Delta}
(\phi_{k_{1}^{(1)},l_{1}^{(1)},
\ldots,k_{m}^{(1)},l_{m}^{(1)}},\ldots,\phi_{k_{1}^{(M)},l_{1}^{(M)},
\ldots,k_{m}^{(M)},l_{m}^{(M)}}) \right ), &&
\end{eqnarray}
where $\density$ is the state density operator acting on the
composite Hilbert space $\mathcal{H}_{\mathcal{Q}}$. Next, we
define to what extent the density operator depends on the
particular joint relative-phase sum
$\phi_{k_{1}^{(1)},l_{1}^{(1)}, \ldots,k_{m}^{(1)},l_{m}^{(1)}}$,
e.g.,
\begin{eqnarray}
\nonumber&&\gamma_{k_{1}^{(1)},l_{1}^{(1)},
\ldots,k_{m}^{(1)},l_{m}^{(1)}}=
\\\nonumber&&
\nonumber |\int_{2\pi} d\phi_{k_{1}^{(1)},l_{1}^{(1)},
\ldots,k_{m1}^{(1)},l_{m}^{(1)}}
e^{-i\phi_{k_{1}^{(1)},l_{1}^{(1)},
\ldots,k_{m}^{(1)},l_{m}^{(1)}}}
\\&&
\mathcal{P}(\phi_{k_{1}^{(1)},l_{1}^{(1)},
\ldots,k_{m}^{(1)},l_{m}^{(1)}},\ldots,\phi_{k_{1}^{(M)},l_{1}^{(M)},
\ldots,k_{m}^{(M)},l_{m}^{(M)}}) |,
\end{eqnarray}
where $\mathcal{P}$ must be expressed in the relative-phase sum
parameter $\phi_{k_{1}^{(1)},l_{1}^{(1)},
\ldots,k_{m}^{(1)},l_{m}^{(1)}}$, but the particular choice of the
remaining $M-1$ linearly independent relative-phase sum parameters
is inconsequential for the absolute value of the integral. The coefficients
$\gamma_{k_{u},l_{u}, \ldots,k_{v},l_{v}}$ (where, here, and in
the following, we will omit the superscript on the indices) are
proportional to the Fourier components of the joint relative-phase
distribution. Now, let us introduce the following index operator
to connect the notation using the subsystem indices, and the
notation using the joint-system index running from 1 to $N_1 N_2
\cdots N_m$:
\begin{eqnarray} \nonumber
&&\Pi(k_{1},l_{1},k_{2},l_{2},\cdots,k_{m},l_{m})=(k_{1}-1)N_{2}\cdots
N_{m}\\\nonumber &&
 +(k_{2}-1)N_{3}\cdots N_{m}+\cdots +(k_{m-1}-1)N_{m}+k_{m} ,\\\nonumber
&&(l_{1}-1)N_{2}\cdots N_{m} +(l_{2}-1)N_{3}\cdots N_{m}\\ && +
\cdots +(l_{m-1}-1)N_{m}+l_{m} .
\end{eqnarray}
Note that the index operator generates two indices based on the
set $\{k_u\}$ and $\{l_u\}$, respectively. Evaluating the Fourier
components, one finds, not surprisingly, that
$\gamma_{k_{1},l_{1},\ldots,k_{m},l_{m}}=2 \pi
|\rho_{\Pi(k_{1},l_{1},\ldots,k_{m},l_{m})}|$. That is, to each
relative-phase sum there is an associated joint-system density
matrix coefficient. We now define an index permutation operator
$\mathrm{P}_{j}$ operating on any function
$f(k_{1},l_{1},\ldots,k_{m},l_{m})$ by
\begin{eqnarray}
\mathrm{P}_{j}f(k_{1},l_{1},\ldots,k_{j},l_{j},\ldots,k_{m},l_{m})
= && \nonumber \\
f(k_{1},l_{1},\ldots,k_{j},l_{j},\ldots,k_{m},l_{m}) && \nonumber
 \\
-f(k_{1},l_{1},\ldots,l_{j},k_{j},\ldots,k_{m},l_{m}) .
\end{eqnarray}
Using this operator we can generalize our earlier results for
bipartite systems \cite{Hosh1,Hosh2}. We form an entanglement
function by summing the absolute difference between pairwise
relative-phase sums. The function is given by
{\begin{widetext} \begin{eqnarray} \nonumber &&\Gamma(\hat{\rho})=
(\mathcal{N}_2 \sum^{N_{1}}_{l_{1}>k_{1}}\sum^{N_{1}}_{k_{1}=1}
\sum^{N_{2}}_{l_{2}>k_{2}}\sum^{N_{2}}_{k_{2}=1}
\sum^{N_{3}}_{k_{3}=l_{3}=1}
\cdots\sum^{N_{m}}_{k_{m}=l_{m}=1}|\mathrm{P}_2|
\rho_{\Pi(k_{1},l_{1},k_{2},l_{2},\ldots,k_{m-1},l_{m-1},k_{m},l_{m})}||^2
+ \ldots
\\\nonumber
&&+\mathcal{N}_2\sum^{N_{m-1}}_{l_{m-1}>k_{m-1}}\sum^{N_{m-1}}_{k_{m-1}=1}
\sum^{N_{m}}_{l_{m}>k_{m}\mathcal{N}_2}\sum^{N_{m}}_{k_{m}=1}
\sum^{N_{1}}_{k_{1}=l_{1}=1}\cdots\sum^{N_{m-2}}_{k_{m-2}=l_{m-2}=1}|\mathrm{P}_m|
\rho_{\Pi(k_{1},l_{1},k_{2},l_{2},\ldots,k_{m-1},l_{m-1},k_{m},l_{m})}||^2\\\nonumber
&&+ \mathcal{N}_3\sum^{N_{1}}_{l_{1}>k_{1}}\sum^{N_{1}}_{k_{1}=1}
\sum^{N_{2}}_{l_{2}>k_{2}}\sum^{N_{2}}_{k_{2}=1}
\sum^{N_{3}}_{l_{3}>k_{3}}\sum^{N_{3}}_{k_{3}=1}
\sum^{N_{4}}_{i_{4}=j_{4}=1}\cdots\sum^{N_{m}}_{k_{m}=l_{m}=1}\{|\mathrm{P}_2|\mathrm{P}_3
|\rho_{\Pi(k_{1},l_{1},k_{2},l_{2},k_{3},l_{3},\ldots,k_{m-1},l_{m-1},k_{m},l_{m})}
||^2|\} + \ldots \\\nonumber &&
+\mathcal{N}_3\sum^{N_{m-2}}_{l_{m-2}>k_{m-2}}\sum^{N_{m-2}}_{k_{m-2}=1}
\cdots \sum^{N_{m}}_{l_{m}>k_{m}}\sum^{N_{m}}_{k_{m}=1}
\sum^{N_{1}}_{k_{1}=l_{1}=1}\cdots\sum^{N_{m-3}}_{i_{m-3}=j_{m-3}=1}\{|\mathrm{P}_{m-1}|\mathrm{P}_{m}
|\rho_{\Pi(k_{1},l_{1},k_{2},l_{2},\ldots,k_{m-1},l_{m-1},k_{m},l_{m})}||^2|\}
\\\nonumber &&
+\ldots+ \\\nonumber &&
+\mathcal{N}_{m-1}\sum^{N_{1}}_{l_{1}>k_{1}}\sum^{N_{1}}_{k_{1}=1}
 \cdots
\sum^{N_{m-1}}_{l_{m-1}>k_{m-1}}\sum^{N_{m-1}}_{k_{m-1}=1}
\sum^{N_{m}}_{k_{m}=l_{m}=1}\{|\mathrm{P}_{2}|\mathrm{P}_{3}|\cdots|\mathrm{P}_{m-1}
|\rho_{\Pi(k_{1},l_{1},k_{2},l_{2},\ldots,k_{m-2},l_{m-2},k_{m-1},l_{m-1},k_{m},l_{m})}||^2\cdots
||| \}\\\nonumber
&&
+\ldots+\mathcal{N}_{m-1}\sum^{N_{2}}_{l_{2}>k_{2}}\sum^{N_{2}}_{k_{2}=1}
\cdots \sum^{N_{m}}_{l_{m}>k_{m}}\sum^{N_{m}}_{k_{m}=1}
\sum^{N_{1}}_{k_{1}=l_{1}=1}\{|\mathrm{P}_{3}|\mathrm{P}_{4}|\cdots\mathrm|{P}_{m}
|\rho_{\Pi(k_{1},l_{1},k_{2},l_{2},k_{3},l_{3},
\ldots,k_{m-2},l_{m-2},k_{m-1},l_{m-1},k_{m},l_{m})}||^2 \cdots
|||\}
\\
&&
+ \mathcal{N}_m \sum^{N_{1}}_{l_{1}>k_{1}}\sum^{N_{1}}_{k_{1}=1}
\cdots \sum^{N_{m}}_{l_{m}>k_{m}}\sum^{N_{m}}_{k_{m}=1}\{
|\mathrm{P}_{2}|\mathrm{P}_{3}|\cdots|\mathrm{P}_{m}|\rho_{\Pi(k_{1},l_{1},k_{2},l_{2},\ldots,k_{m-1},l_{m-1},k_{m},l_{m})}||^2
\cdots||| \}
 )^{\frac{1}{2}} .
\label{eq:entanglement measure}
\end{eqnarray}
\end{widetext}}
This is our central equation. It looks messy, but has a rather
logical inner structure. The factors $\mathcal{N}_u$ are
normalization factors, and they should not be confused with the
space dimensions $N_u$. The first sums, where two of them are
written explicitly (with normalization factors $\mathcal{N}_2$) on
the right hand side of (\ref{eq:entanglement measure}), only
contributes the
respective subsystem's bipartite entanglement. There are $\left(%
\begin{array}{c}
  m \\
  2 \\
\end{array}%
\right)=m(m-1)/2$ ways to select two systems out of $m$ without
respect to ordering. The two terms explicitly written above sums
the bipartite entanglement contribution between systems
$\mathcal{Q}_{1},\mathcal{Q}_{2}$ and
$\mathcal{Q}_{m-1},\mathcal{Q}_{m}$, respectively. For the systems
$\{\mathcal{Q}_{u},\mathcal{Q}_{v}\}$, there are
$N_u(N_u-1)N_v(N_v-1)/4$ ways to select one each of the relative
phases of system $\mathcal{Q}_{u}$ and $\mathcal{Q}_{v}$. Because
the other system's coefficients can be chosen arbitrarily among
the diagonals, there are $\Pi_{j=1}^m N_j/ (N_u N_v)$ number of
relative-phase sums and differences involving $k_u,l_u,k_v$ and
$l_v$. Our permutation operator subtracts the relative-phase
difference from the relative-phase sum, so by including all
bipartite combinations, the bipartite entanglement of the joint
system is taken care of. Next, we add the tripartite entanglement
(contained in the sums with normalization factors
$\mathcal{N}_3$). There are $\left(
\begin{array}{c}
  m \\
  3 \\
\end{array}
\right)$ tripartite combinations, and for every choice
$\{\mathcal{Q}_{u},\mathcal{Q}_{v},\mathcal{Q}_{w}\}$, where
$u<v<w$, there are $N_u(N_u-1)N_v(N_v-1)N_w(N_w-1)/8$ combination
of system relative phases. For each combination, we can sum all
three relative phases, sum the first two and subtract the third,
etc. To form differences of all combinations, we use both the
permutation operators $\mathrm{P}_{v}$ and $\mathrm{P}_{w}$.
Hence, we get $4=2^{3-1}$ contributions within the first curly
bracket in (\ref{eq:entanglement measure}), above. For each
choice, the other systems indices can be chosen in $\Pi_{j=1}^m
N_j/ (N_u N_v N_w)$ different ways. For the quadripartite
contribution we proceed in the same way. For every choice
$\{\mathcal{Q}_{u},\mathcal{Q}_{v},\mathcal{Q}_{w},\mathcal{Q}_{z}\}$,
where $u<v<w<z$, we use the permutation operators
$\mathrm{P}_{v}$, $\mathrm{P}_{w}$, and $\mathrm{P}_{z}$. We get
$8=2^{4-1}$ contributions inside the corresponding curly brackets,
each being a sum of $\Pi_{j=1}^m N_j/ (N_u N_v N_w N_z)$ terms.
The sum proceed in this fashion until the $m$-partite entanglement
contributions are to be added. There is only one way ($\left(
\begin{array}{c}
  m \\
  m \\
\end{array}
\right)=1$) to chose all subsystems, and we insert $m$ as index in
our permutation operator. we use the permutation operators
$\mathrm{P}_{2}$, $\mathrm{P}_{3}$, \ldots , $\mathrm{P}_{m}$. (We
do not permute $k_1$ and $l_1$.) In all, we get $2^{m-1}$ terms
inside the curly brackets of the last sum in (\ref{eq:entanglement
measure}), above. These terms represent all the possible
relative-phase sums and differences between all the $m$-systems,
so there are no further terms.

From our definitions, it is clear that for any product state
\beq\rho_{\Pi(k_{1},l_{1},k_{2},l_{2},\ldots,k_{m},l_{m})} =
\rho_{k_{1},l_{1}} \rho_{k_{2},l_{2}}\cdots\rho_{k_{m},l_{m}}
,\eeq where $ \rho_{k_{u},l_{u}}$ is the indicated density matrix
coefficient of system $u$. In this case, one gets
$\mathrm{P}_u|\rho_{\Pi(k_{1},l_{1},\ldots,k_{m},l_{m})}|=0$ for
any $u$ and any set of indices $k_{1},l_{1},\ldots,k_{m},l_{m}$.
Hence, our entanglement function $\g(\oprho)=0$ for any tensor
product of $m$ density operators. For entangled states, the
function is not invariant to local unitary transformations. In
analogy with our definitions for bipartite states, {\em we define our
measure of entanglement $\gmax $, where sup refers to the supremum
of $\g$ under all possible local unitary transformations}.


Let us now write out and use (\ref{eq:entanglement measure}) in a
few explicit cases. The degree of entanglement for a
$\mathcal{H}_{\mathcal{Q}_{1}}\otimes\mathcal{H}_{\mathcal{Q}_{2}}$
bipartite states is given by
\begin{eqnarray}
\Gamma(\hat{\rho})&=& \nonumber (\mathcal{N}_2
\sum^{N_{1}}_{l_{1}>k_{1}}\sum^{N_{1}}_{k_{1}=1}
\sum^{N_{2}}_{l_{2}>k_{2}}\sum^{N_{2}}_{k_{2}=1}
||\rho_{(k_{1}-1)N_{2}+k_{2},(l_{1}-1)N_{2}+l_{2}}|\\&&
-|\rho_{(k_{1}-1)N_{2}+l_{2},(l_{1}-1)N_{2}+k_{2}}||^{2}_{\mathcal{Q}_{1}\mathcal{Q}_{2}}
)^{\frac{1}{2}}.
\end{eqnarray}
This special case has already been discussed in detail in
\cite{Hosh1,Hosh2}, and we have shown that the equation coincides
with the concurrence \cite{Wootters} for pure bipartite states in
$2 \otimes 2$ (provided that one sets $\mathcal{N}_2 = 2$) and
with generalized concurrence measures in $2 \otimes 3$ dimensions
\cite{Rungta,Albeverio,Audenaert}.

Note that our measure sums all the state's entanglement. That is,
although, e.g., a state's bipartite entanglement between
subsystems $\mathcal{Q}_{1}$ and $\mathcal{Q}_{2}$ cannot be used
simultaneously neither with its bipartite entanglement between
subsystems $\mathcal{Q}_{1}$ and $\mathcal{Q}_{3}$, nor, e.g.,
with its tripartite entanglement between subsystems
$\mathcal{Q}_{1}$, $\mathcal{Q}_{2}$, and $\mathcal{Q}_{3}$, all
contributions are added in our measure. That is, our measure
characterizes the entanglement contained in a state, but in
general the measure exceeds the usable entanglement. However, by
looking at the various terms in the sum, the usable entanglement
can be extracted as the measure is composed of sub-sums containing
the bipartite $\mathcal{Q}_{1}$ and $\mathcal{Q}_{2}$
entanglement, the bipartite $\mathcal{Q}_{1}$ and
$\mathcal{Q}_{3}$ entanglement, the tripartite $\mathcal{Q}_{1}$,
$\mathcal{Q}_{2}$, and $\mathcal{Q}_{3}$ entanglement, etc., as
can explicitly be seen in (\ref{eq: explicit example}), below.
Also note that our measure sums the possible cooperative
entanglement. That is, if some subsystems are ignored, or the
information contained in a subsystem is lost, then the ensuing
state's entanglement is in general lower than what our measure
predicts. We shall give a concrete example of this in Sec.
\ref{sec: beyond}, below.


\section{Tripartite entanglement}

The degree of entanglement for a
$\mathcal{H}_{\mathcal{Q}_{1}}\otimes\mathcal{H}_{\mathcal{Q}_{2}}
\otimes\mathcal{H}_{\mathcal{Q}_{3}}$ tripartite state is given by
\begin{widetext}
\begin{eqnarray}\label{eq: tripartite states } \nonumber
\Gamma(\hat{\rho})&=& (\mathcal{N}_2
[\sum^{N_{1}}_{l_{1}>k_{1}}\sum^{N_{1}}_{k_{1}=1}
\sum^{N_{2}}_{l_{2}>k_{2}}\sum^{N_{2}}_{k_{2}=1}
\sum^{N_{3}}_{k_{3}=l_{3}=1}||
\rho_{\Pi(k_{1},l_{1},k_{2},l_{2},k_{3},l_{3})}|
-|\rho_{\Pi(k_{1},l_{1},l_{2},k_{2},k_{3},l_{3})}|
|^{2}_{\mathcal{Q}_{1}\mathcal{Q}_{2}}\\\nonumber && +
\sum^{N_{1}}_{l_{1}>k_{1}}\sum^{N_{1}}_{k_{1}=1}
\sum^{N_{3}}_{l_{3}>k_{3}}\sum^{N_{3}}_{k_{3}=1}\sum^{N_{2}}_{k_{2}=l_{2}=1}||
 \rho_{\Pi(k_{1},l_{1},k_{2},l_{2},k_{3},l_{3})}|
-| \rho_{\Pi(k_{1},l_{1},k_{2},l_{2},l_{3},k_{3})}|
|^{2}_{\mathcal{Q}_{1}\mathcal{Q}_{3}}
\\&&\nonumber
+\sum^{N_{2}}_{l_{2}>k_{2}}\sum^{N_{2}}_{k_{2}=1}
\sum^{N_{3}}_{l_{3}>k_{3}}\sum^{N_{3}}_{k_{3}=1}
\sum^{N_{1}}_{k_{1}=l_{1}=1}||
\rho_{\Pi(k_{1},l_{1},k_{2},l_{2},k_{3},l_{3})}| -|
\rho_{\Pi(k_{1},l_{1},k_{2},l_{2},l_{3},k_{3})}|
|^{2}_{\mathcal{Q}_{2}\mathcal{Q}_{3}}]
\\&&\nonumber
+\mathcal{N}_3\sum^{N_{1}}_{l_{1}>k_{1}}\sum^{N_{1}}_{k_{1}=1}
\sum^{N_{2}}_{l_{2}>k_{2}}\sum^{N_{2}}_{k_{2}=1}
\sum^{N_{3}}_{l_{3}>k_{3}}\sum^{N_{3}}_{k_{3}=1}
 \{ |||\rho_{\Pi(k_{1},l_{1},k_{2},l_{2},k_{3},l_{3})}|
-|\rho_{\Pi(k_{1},l_{1},k_{2},l_{2},l_{3},k_{3})}|
|^{2}_{\mathcal{Q}_{1}\mathcal{Q}_{2}\mathcal{Q}_{3}}
\\&&
-|| \rho_{\Pi(k_{1},l_{1},l_{2},k_{2},k_{3},l_{3})}| -|
\rho_{\Pi(k_{1},l_{1},l_{2},k_{2},l_{3},k_{3})}|
|^{2}_{\mathcal{Q}_{1}\mathcal{Q}_{2}\mathcal{Q}_{3}}|
\})^{\frac{1}{2}} .
\label{eq: explicit example}
\end{eqnarray}
\end{widetext}


Let us now give two concrete examples of this measure for some
three-qubit states. In the three-qubit space there exist two
classes of states, inequivalent under local operations and
classical communication (LOCC), called $\ket{\Psi_{{\rm GHZ}}}$
and $\ket{\Psi_{{\rm W}}}$ states. They are, e.g., $
\ket{\Psi_{{\rm GHZ}}}=(\ket{000}+\ket{111})/\sqrt{2}$ and $
\ket{\Psi_{{\rm W}}}=(\ket{001}+\ket{010}+\ket{100})/\sqrt{3}$.
For these states, we have
$$
\Gamma(\hat{\rho}_{{\rm GHZ}})= \left(\mathcal{N}_3
\left|\rho_{1,8}\right|^{2}_{\mathcal{Q}_{1}\mathcal{Q}_{2}\mathcal{Q}_{3}}\right)^{\frac{1}{2}}
=\sqrt{\frac{\mathcal{N}_3}{4}},
$$
 and
\begin{eqnarray}
\Gamma(\hat{\rho}_{{\rm W}}) &=&\nonumber
 \nonumber\left(\mathcal{N}_2(
\left|\rho_{3,5}\right|^{2}_{\mathcal{Q}_{1}\mathcal{Q}_{2}} +
\left|\rho_{2,5}\right|^{2}_{\mathcal{Q}_{1}\mathcal{Q}_{3}}
+\left|\rho_{2,3}\right|^{2}_{\mathcal{Q}_{2}\mathcal{Q}_{3}})\right)^{\frac{1}{2}}
\\\nonumber
&=&\sqrt{\frac{\mathcal{N}_2}{3}}.
\end{eqnarray}
Here, we see that the normalization factors must be retained (or,
possibly be chosen with particular relative weights) in order for
the entanglement measure to make sense for general states. Because
the GHZ- and the W-states belong to different equivalence classes,
their relative entanglement weights are not obvious. This issue is
tied to the, still open, question about minimum reversible
entanglement generating sets
\cite{Dur99,Dur00,Bennett01,Acin01,Galvao}.


\section{Beyond tree-partite qubit entanglement}
\label{sec: beyond}

Next, let us look at an interesting four-qubit state
$\ket{\Psi_1}=(\ket{0,0,0,1}+\ket{0,1,0,0}+\ket{1,0,1,0}+\ket{1,1,1,1})/2
.$ Our measure of entanglement of this state is
\begin{eqnarray} \nonumber \Gamma(\Psi_1)&=&(\mathcal{
N}_{2}(|\rho_{2,5}|^{2}_{\mathcal{Q}_{2}\mathcal{Q}_{4}}+|\rho_{11,16}|^{2}_{\mathcal{Q}_{2}\mathcal{Q}_{4}})\\\nonumber&&+\mathcal{
N}_{3}(|\rho_{2,16}|^{2}_{\mathcal{Q}_{1}\mathcal{Q}_{2}\mathcal{Q}_{3}}
+|\rho_{5,11}|^{2}_{\mathcal{Q}_{1}\mathcal{Q}_{2}\mathcal{Q}_{3}}\\\nonumber&&
+|\rho_{2,11}|^{2}_{\mathcal{Q}_{1}\mathcal{Q}_{3}\mathcal{Q}_{4}}
+|\rho_{5,16}|^{2}_{\mathcal{Q}_{1}\mathcal{Q}_{3}\mathcal{Q}_{4}}))^{\frac{1}{2}}\\\nonumber
&=& \left (\frac{\mathcal{N}_{2}}{8} + \frac{\mathcal{N}_{3}}{4}
\right )^{\frac{1}{2}}.
\end{eqnarray}
The state has both bipartite and tripartite entanglement. In order
to use the bipartite entanglement the parties possessing the
different qubit subsystems must cooperate. If, e.g., qubit 1 and 3
are measured in the standard basis, the result is either two zeros
or two ones. If this result is communicated to the parties holding
qubit 2 and 4, (that is, we perform a LOCC, optimal for bringing
out the bipartite entanglement) the remaining two-qubit state will
be in (a known) pure EPR-state. If, on the other hand, if we
simply trace out subsystems $\mathcal{Q}_{1}$ and
$\mathcal{Q}_{3}$, (or measure qubit 1 and 3 but keep the result
secret), then the remaining state is in an an equal mixture of the
EPR-states, and this state is directly separable. This means that
if the different parties do not cooperate, the state's bipartite
entanglement in subspace 1 and 3 vanishes.

The entanglement of the state
$\ket{\Psi_2}=(\ket{0,1,1,0}+\ket{1,0,0,1}+\ket{0,1,1,1}+\ket{1,0,0,0})/2
,$ on the other hand, is given by $$\Gamma(\Psi_2)=
(\mathcal{N}_{3}|\rho_{7,9}|^{2}_{\mathcal{Q}_{1}\mathcal{Q}_{2}\mathcal{Q}_{3}})^{\frac{1}{2}}
= \sqrt{\frac{\mathcal{N}_{3}}{4}} .$$ That is, the state has only
tripartite entanglement and no bipartite entanglement. To arrive
at the result, we note that a unitary transformation $\hat{U}_{4}$
local to $\mathcal{Q}_{4}$ can transform the state into, e.g.,
$\hat{U}_{4}\ket{\Psi_2}=(\ket{0,1,1}+\ket{1,0,0}\otimes\ket{0})/\sqrt{2}$
for which one finds the supremum of $\g$. In this case, the
state's entanglement is the same whether or not the person in
possession of qubit 4 cooperates or not.


As a last example, consider a a $m$-qubit density operator given
by a mixture of the two orthogonal, so called, $m$-Cat states $$
\ket{\Psi_{{\rm
Cat}}}=\frac{1}{\sqrt{2}}(\ket{0_{1},0_{2},\ldots,0_{m}}+\ket{1_{1},1_{2},\ldots,1_{m}})$$.
 Then, our degree of
entanglement gives \begin{eqnarray} \nonumber \Gamma(\Psi_{{\rm
Cat}}) &=&\left(\mathcal{N}_m
\left|\rho_{1,2^{m}}\right|^{2}_{\mathcal{Q}_{1}\mathcal{Q}_{2}\cdots\mathcal{Q}_{m}}\right)^{\frac{1}{2}}
\\\nonumber &=&
 \left(\frac{\mathcal{N}_m}{4}\right)^{\frac{1}{2}}.
\end{eqnarray}

\section{Conclusions}

In conclusion, we have proposed an entanglement measure for pure
multipartite quantum states. The measure directly detects product
states (it is zero for such states), and quantifies the
entanglement of any pure state up to the bipartite, tripartite,
\ldots, m-partite normalization coefficients. Since it is not
possible to use, nor convert, the entanglement in states with
incompatible entanglement classes such as GHZ- and W-states into
each other, it may not be meaningful to specify the coefficients
relative to each other. Rather, from an operational point of view,
it seems more meaningful to specify each type of entanglement
separately, e.g. in a system composed of four subsystems
$\mathcal{Q}_{1}$, $\mathcal{Q}_{2}$, $\mathcal{Q}_{3}$, and
$\mathcal{Q}_{4}$, it is meaningful to discuss, separately, the
bipartite entanglement between, e.g., systems $\mathcal{Q}_{1}$
and $\mathcal{Q}_{2}$, and $\mathcal{Q}_{1}$ and
$\mathcal{Q}_{4}$. We do not see how the bipartite entanglement,
in an operational sense, could (or should) be compared to, e.g.,
the tripartite entanglement between $\mathcal{Q}_{1}$,
$\mathcal{Q}_{2}$, and $\mathcal{Q}_{4}$. Our measure sums all
contributions to quantify the state's entire entanglement, but, as
just indicated, from an operational viewpoint, it is rather the
the sum's various contributions that have a well defined
operational meaning. This is in contrast to, e.g., Partovi's
measure \cite{Partovi}, that is a minimum entropic distance
measure between the state and a separable state with the same
statistical marginal distributions. In this sense, Hossein
Partovi's measure is a better measure of the ``quantumness'' of
the state, while our measure emphasize the state's usefulness as a
quantum information carrier.

\begin{acknowledgments}
We wish to thank Professor L. L. S\'{a}nchez-Soto for inspiration
and useful discussions. This work was supported by the Swedish
Research Council (VR) and the Swedish Foundation for Strategic
Research (SSF).
\end{acknowledgments}


\begin{thebibliography}{99}
\bibitem{Lewen00} M. Lewenstein, D. Bru\ss, J. I. Cirac, B. Kraus, M. Ku\'{s}, J. Samsonowicz, A. Sanpera, and R. Tarrach, J. Mod. Opt. {\bf 47}, 2841 (2000).
\bibitem{Dur99}W. D\"{u}r, J. I. Cirac, and R. Tarrach, \prl {\bf 83}, 3562 (1999).
\bibitem{Vedral97} V. Vedral, M. B. Plenio, M. A. Rippin, and P.~L.~Knight, \prl {\bf 78}, 2275 (1997).
\bibitem{Vedral98} V. Vedral, M. B. Plenio, K. Jacobs, and P.~L.~Knight, \pra  {\bf 58}, 883 (1998).
\bibitem{Werner89} R. F. Werner, \pra  {\bf 40}, 4277 (1989).
\bibitem{Wootters} W. K. Wootters, Phys. Rev. Lett. {\bf 80}, 2245 (1998).
\bibitem{Rungta} P. Rungta, V. Bu\v{z}ek, C. M. Caves, M. Hillery, and G.~J.~Milburn, Phys. Rev. A {\bf 64}, 042315 (2001).
\bibitem{Albeverio} S. Albeverio and S. M. Fei, J. Opt. B: Quantum Semiclass. Opt. {\bf
3}, 223 (2001).
\bibitem{Audenaert} K. Audenaert, F. Verstraete, and B. De Moor,
Phys. Rev. A {\bf 64}, 012316 (2001).
\bibitem{Gerjuoy} E. Gerjuoy, \pra {\bf 67}, 052308 (2003).
\bibitem{Plenio00} M. B. Plenio and  V. Vedral, J. Phys. A: Math. Gen. {\bf 34}, 6997 (2001).
\bibitem{Bennett01} C. H. Bennett, S. Popescu, D. Rohrlich, J. Smolin, and  A. V. Thapliyal, \pra {\bf 63}, 012307 (2001).
\bibitem{Dur00} W. D\"{u}r, G. Vidal, and J. I. Cirac, \pra  {\bf 62}, 062314 (2000).
\bibitem{Eisert01} J. Eisert and H. J. Briegel, \pra {\bf 63}, 022306 (2000).
\bibitem{Verst03} F. Verstraete, J. Dehaene, and B. De Moor, \pra {\bf 68},
012103 (2003).
\bibitem{Eckert02} K. Eckert, O. G\"{u}hne, F. Hulpke, P. Hyllus, J. Korbicz, J.~Mompart, D. Bru\ss, M. Lewenstein, and  A. Sanpera, in {\em Quantum Information Processing}, edited by G. Leuchs and T. Beth (Wiley-VCH, Berlin 2003).
\bibitem{Hor00} M. Horodecki, P. Horodecki, and R. Horodecki, Phys. Lett. A {\bf 283}, 1 (2001).
\bibitem{Acin01} A. Ac\'{i}n, D. Bru\ss, M. Lewenstein, and A. Sanpera,  \prl {\bf 87}, 040401 (2001).
\bibitem{Partovi} M. H. Partovi, \prl  {\bf 92}, 077904 (2004).
\bibitem{Soto02} L. L. S\'{a}nchez-Soto, J. Delgado, A. B. Klimov, and G.~Bj\"{o}rk, \pra {\bf 66}, 042112 (2002).
\bibitem{Hosh1} H. Heydari, G.~Bj\"{o}rk, and L. L. S\'{a}nchez-Soto, \pra {\bf 68}, 062314 (2003).
\bibitem{Hosh2} H. Heydari and  G.~Bj\"{o}rk, e-print quant-ph/0401128 v1.
\bibitem{Galvao} E. F. Galv\~ao, M. B. Plenio, and S. Virmani, J. Phys. A: Math. Gen. {\bf 33}, 8809 (2000).
\end{thebibliography}
\end{document}